%% file: project.tex
\documentclass[final,authoryear,5p,times, twocolumn]{article}

\usepackage{lineno}
\usepackage{graphicx}
\usepackage{color}
\usepackage[dvipsnames]{xcolor}
\usepackage{amssymb}
\usepackage{amsmath}
\usepackage{bigints}
\usepackage[bookmarksdepth=3]{hyperref}
\usepackage{lineno}
\usepackage{mathtools}
\usepackage{blkarray}
\usepackage[capposition=bottom]{floatrow}
\usepackage{varwidth}
\usepackage{pifont}
\graphicspath{{}}
\usepackage{booktabs}
\usepackage{pgf,tikz}
\usepackage{mathrsfs}
\usetikzlibrary{arrows}
\usepackage{authblk}
\hypersetup{
    colorlinks=true,
    citecolor=red,
    linkcolor=blue,
    filecolor=blue,      
    urlcolor=red,}

\title{Algebraic 3D Graphic Statics: reciprocal constructions}
      
\date{}

\author[$a$, $b$]{M\'{a}rton Hablicsek}

\author[$a$]{Masoud Akbarzadeh}

\author[$a$]{Yi Guo}

\affil[$a$]{Polyhedral Structures Laboratory, School of Design, University of Pennsylvania, Philadelphia, USA}

\affil[$b$]{Department of Mathematics, Centre of Symmetry and Deformation, University of Copenhagen, Denmark}

\begin{document}
\twocolumn[
  \begin{@twocolumnfalse}
    \maketitle

\begin{abstract}
The recently developed 3D graphic statics (3DGS) lacks a rigorous mathematical definition relating the geometrical and topological properties of the reciprocal polyhedral diagrams as well as a precise method for the geometric construction of these diagrams. This paper provides a fundamental algebraic formulation for 3DGS by developing equilibrium equations around the edges of the primal diagram and satisfying the equations by the closeness of the polygons constructed by the edges of the corresponding faces in the dual/reciprocal diagram. The research provides multiple numerical methods for solving the equilibrium equations and explains the advantage of using each technique. The approach of this paper can be used for compression-and-tension combined form-finding and analysis as it allows constructing both the form and force diagram based on the interpretation of the input diagram. Besides, the paper expands on the geometric/static degrees of (in)determinacies of the diagrams using the algebraic formulation and shows how these properties can be used for the constrained manipulation of the polyhedrons in an interactive environment without breaking the reciprocity between the two.


\vspace{0.5cm}

\end{abstract}
  \end{@twocolumnfalse}
]


\input{int.tex}
\input{theoritical.tex}

\input{implementation.tex}
\input{application.tex}

\input{conclusion.tex}

\bibliographystyle{plain}
\bibliography{3DGS_refs}

\end{document}

%% file: int.tex
\section{Introduction}
 
In graphic statics, the geometry of the structure and its equilibrium are represented by the \textit{form} and the \textit{force} diagrams where the length of the members, the location of the supports and the applied loads are represented by the former, and the equilibrium and the magnitude of the forces are represented by the latter. These two diagrams are \textit{reciprocal}, i.e. topologically dual and geometrically dependent \cite{maxwell1864}. In fact, the methods of graphic statics is essentially the geometric construction of these two reciprocal diagrams for various geometries, loading cases, and boundary conditions. 

\subsection{Reciprocal diagrams and their constructions}

In 2D graphic statics, as it was developed and practiced in the late nineteenth century, the construction of the reciprocal diagrams was a step-by-step geometric construction \cite{culmann1864, Bow1873, cremona1890, wolfe1921}. This procedural approach is quite cumbersome and lengthy for the structures with multitudes of members, and any design iteration requires a new construction process. This slow workflow could be the reason for the shift towards the development of the numerical methods at the end of the nineteenth century.

 Graphic statics in combination with computational methods result in innovative design tools allowing the exploration of the realm of unique, sophisticated, yet efficient structural solutions. Using computational methods can significantly accelerate the construction of the reciprocal diagrams and exploit the explicit relationship between the form of a structure and its geometric equilibrium of forces in an interactive environment \cite{block2009, Rippmann2012}.

The topological and geometrical relationships between the reciprocal diagrams of 2D graphics statics (2DGS) have recently been formulated as algebraically--constrained equations whose numerical solutions allows the direct construction of the diagrams in an interactive environment 
\cite{VANMELE2014104, ALIC201726}. Besides, the algebraic formulation of the graphic statics is a rigorous approach providing an in-depth understanding of some essential properties such as the geometric/static degrees of indeterminacies of both form and force diagrams. These parameters can be used interactively to manipulate the geometry of these diagrams without breaking the reciprocity between them.   

\subsection{Problem Statement and Objectives}
3D Graphical statics is a recent development of graphic statics in three dimensions based on a historical proposition by \cite{rankine1864} and \cite{maxwell1864} \cite{Akbarzadeh2015, akbarzadeh2016_phd, Williams2016121, McRobie2017, LEE201811}. In 3DGS, the form and the force diagrams are polyhedral diagrams; the equilibrium of each node of the form with its applied loads/members is represented by a closed force polyhedron whose faces are perpendicular to the loads/members of the node. The area of each face of the force polyhedron represents the magnitude of the force in the corresponding member of the node.

Similar to 2DGS, the geometric construction of the reciprocal polyhedral diagrams is the most crucial step in using 3DGS methods. \cite{akbarzadeh20153d} explained a step-by-step procedural approach to construct both form and force diagrams of 3DGS for a given boundary conditions and loading scenario with the similar drawbacks of the procedural 2DGS methods. Additionally, \cite{Akbarzadeh2014} suggested a computational implementation based on iterative geometric construction to find reciprocal forms for a given group of closed, convex polyhedral cells. Although the proposed method is quite robust in generating the reciprocal diagrams, the precise control of the edge lengths of the members of the diagrams is quite challenging. Moreover, the method cannot construct the reciprocals for complex/self-intersecting polyhedrons representing the systems with both tension and compression members. Besides, any manipulation introduced by the user in the geometry of the form or force diagram breaks the reciprocity and requires a new iterative computation. In another research, \cite{Konstantatou} suggested the projection of the polyhedral system to the fourth dimension and projecting it back to the third dimension by using paraboloid of revolution that might be relatively counter-intuitive for the users with limited experience with geometric constructions in 3D space. 

In fact, in all mentioned methods, there is a lack of a proper mathematical/algebraic formulation for the reciprocal polyhedral diagrams limiting the interactive implementation of 3DGS. Thus, the primary objective of this paper is to provide an algebraic formulation to relate the reciprocal diagrams and a comprehensive approach to construct and manipulate the reciprocal polyhedrons for compression/tension-only systems as well as the systems with both tension and compression forces.

\subsection{Paper outlines and contributions}
Section \ref{sec:theory} of this paper explains the theoretical framework of the research in the following order:  the essential properties of the form and force diagrams including the nodal, global, and self--stressed polyhedrons (\S \ref{sec:form_force}); the topological properties as well as the incidence matrices to describe the connectivity of the components of the primal and the dual diagrams (\S \ref{sec:topo_geo}, \ref{sec:connect}); the algebraic constraints between two reciprocal diagrams and the process of developing the equilibrium equations to find the lengths of the edges of the dual diagram (\S \ref{sec:algreccon}); and the solution space for the equilibrium equations and the methods to construct the geometry of the dual (\S \ref{sec:solutions}, \ref{sec:condual}).
The algebraic approach of this research can construct the reciprocal diagram for both form and force diagram as the primal input. Therefore, the Section \ref{sec:theory} also explains the procedures for the primal to be considered as a form or force diagram in the approach (\S \ref{sec:prime_force}, \ref{sec:primal_as_form}), and expands on the geometric and static degrees of (in)determinacies of the systems based on the properties of the equilibrium matrix (\S \ref{sec:dsi}).     

Section \ref{sec:comp_set} explains the computational implementation of the algebraic formulation of 3DGS and provides three different numerical methods for solving the equilibrium equations. In Section \ref{sec:appl}, the form finding, analysis, and constrained polyhedral manipulation applications of the presented method are explained and finally the limitations, and the future research directions for this research are listed in Section \ref{sec:conc_disc}. 

\subsection{Nomenclatures}
We denote the algebra objects of this paper as follows; matrices are denoted by bold capital letters (e.g. $\textbf{A}$); vectors are denoted by lowercase, bold letters (e.g., $\textbf{v}$), except the user input vectors which are represented by Greek letters (e.g., $\lambda$); the topological data of the primal diagram are described by italic letters (e.g., $f$); and the data corresponding to the dual and reciprocal diagram are represented by italic letters with a $\dagger$ sign (e.g., $f^\dagger$). Table \ref{table:tab1} encompasses all the notation used in the paper.

\begin{table}[ht!]

\caption{Nomenclature for the symbols used in this paper and their corresponding descriptions.}
\resizebox{\columnwidth}{!}{%
\centering
\begin{tabular}{ll}
\hline
\textit{Topology}       & \textit{Description}                                                                                       \\ \hline
$\Gamma$                        & primal diagram                                                                                    \\
$\Gamma^\dagger$                & dual, reciprocal diagram                                                                          \\
$v$                             & \# of vertices of $\Gamma$                                                                    \\
$e$                             & \# of edges of $\Gamma$                                                                       \\
$f$                             & \# of faces of $\Gamma$                                                                       \\
$c$                             & \# of cells of $\Gamma$                                                                       \\
$v^\dagger$                     & \# of vertices of $\Gamma^\dagger$                                                            \\
$e^\dagger$                     & \# of edges of $\Gamma^\dagger$                                                               \\
$f^\dagger$                     & \# of faces of $\Gamma^\dagger$                                                               \\
$c^\dagger$                     & \# of cells of $\Gamma^\dagger$                                                               \\ \hline
\textit{Matrices}               &                                                       \\ \hline                                            
$\textbf{C}_{e\times v}$        & edge-vertex connectivity matrix of $\Gamma$                                                       \\
$\textbf{C}_{e\times f}$        & edge-face connectivity matrix of $\Gamma$                                                         \\
$\textbf{C}_{f\times c}$        & face-cell connectivity matrix of $\Gamma$                                                         \\
$\textbf{A}$                    & equilibrium matrix                                                                                \\

$\textbf{A}^{+}$              & Moore-Penrose inverse of $\textbf{A}$                                                             \\
$\textbf{A}^{rref}$             & Reduced Row Echelon form of $\textbf{A}$                                                          \\
$\textbf{A}^{rref}_{r\times f}$ & obtained by deleting all zero rows of $\textbf{A}^{rref}$                             \\
$\textbf{N}_x$                  & diagonal matrix of the $x$-coords of $\mathrm{\hat{\textbf{n}}}_i$ \\
$\textbf{N}_y$                  & diagonal matrix of the $y$-coords of the $\mathrm{\hat{\textbf{n}}}_i$ \\
$\textbf{N}_z$                  & diagonal matrix of the $z$-coords of the $\mathrm{\hat{\textbf{n}}}_i$ \\
$\textbf{L}^\dagger$            & Laplacian of $\textbf{C}_{f\times c}$                           \\ \hline
\textit{Vectors}                &                                                       \\ \hline             
$\mathrm{\hat{\textbf{n}}}_i$   & unit normal vector of face $f_i$                                              \\
$\textbf{x}$                    & $x$-coords of $v$                                                        \\
$\textbf{y}$                    & $y$-coords of $v$                                                        \\
$\textbf{z}$                    & $z$-coords of $v$                                                        \\
$\textbf{u}$                    & $x$-coord differences of $v$                                             \\
$\textbf{v}$                    & $y$-coord differences of $v$                                             \\
$\textbf{w}$                    & $z$-coord differences of $v$                                             \\
$\textbf{x}^\dagger$            & $x$-coords of $v^\dagger$                                                \\
$\textbf{y}^\dagger$            & $y$-coords of $v^\dagger$                                                \\
$\textbf{z}^\dagger$            & $z$-coords of $v^\dagger$                                                \\
$\textbf{u}^\dagger$            & $x$-coord differences of $v^\dagger$                                     \\
$\textbf{v}^\dagger$            & $y$-coord differences of $v^\dagger$                                    \\
$\textbf{w}^\dagger$            & $z$-coord differences of $v^\dagger$                                    \\
$\textbf{q}$                    & solution of the equilibrium equations                                                       \\ \hline
\textit{Parameters}          &                                                         \\ \hline    
$\sigma$                        & parameter fixing the location of a vertex of $\Gamma^\dagger$                                        \\
$\xi$                           & parameter for the Moore-Penrose inverse method                                   \\
$\zeta$                         & parameter for RREF method                                          \\
$\lambda$                       & parameter for the Linear programming method                                                       \\ \hline

\textit{Other}  &\\
\hline
$r$                             & rank of $\textbf{A}$                                                                              \\
$\psi_{e_i}$ & indicator of the type of internal forces of $e_i^\dagger$
\\ \hline  
\end{tabular}
}
\label{table:tab1}
\end{table}

%% file: theoritical.tex
\section{Theoretical Framework}
\label{sec:theory}
In this section, we briefly explain the properties of the reciprocal polyhedral diagrams of the 3DGS and set a foundation to describe the algebraic approach to construct these diagrams using a simple example.    

\subsection{Form and force diagrams as groups of polyhedral cells}
\label{sec:form_force}
In the context of 3DGS, both form and force diagrams consists of polyhedral cells in which there is an external polyhedron including all the external faces, and the rest of the cells are inside the external polyhedron. Each edge shares an identical vertex with its adjacent edges, and similarly, each face shares an identical edge with its adjacent faces and finally, each cell shares an identical face with its adjacent cells.

\begin{figure}
  \includegraphics[width=.8\columnwidth]{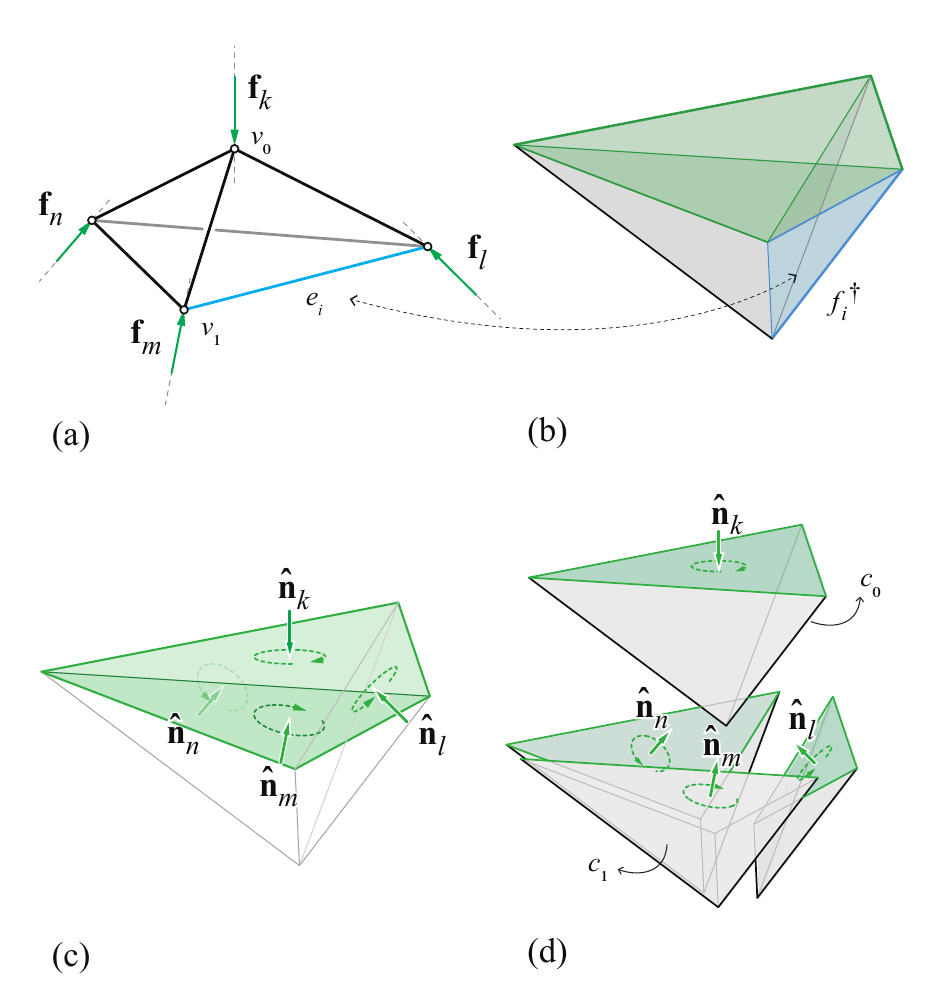}
  \caption{
 (a) A polyhedral structure with an applied load and reaction forces at the support and (b) its corresponding force diagram consisting of $10$ faces and $5$ polyhedral cells; (c) the global force polyhedron (GFP) with the direction of its faces toward inside of the cell; and (d) the faces of GFP constructs nodal force polyhedrons (NFP) whose directions are inherited from the faces of the GFP (three cells toward outside (e.g. $c_1$) and one toward inside ($c_0$).}
 
  \label{fig:det_gfp}
\end{figure}

\subsubsection*{Global and nodal force polyhedrons}
The force diagram of 3DGS consists of closed polyhedral cells that can be decomposed into the following: a \textit{global cell} or \textit{global force polyhedron} (GFP), and \textit{nodal cells} or \textit{nodal force polyhedrons} (NFP) \cite{akbarzadeh2016_phd, Lee2016}. A GFP represents the static equilibrium of externally applied loads and reaction forces regardless of the geometry/topology of the structure. Each NFP represents the equilibrium of forces coming together at that node in the form diagram. Similar to the form diagram, each cell in a group of cells can be chosen as the GFP; if GFP is the external polyhedron, the force diagram can represent a compression/tension--only structural form. While, if GFP is any other cell except the external cell, the force diagram will represent the equilibrium of a force configuration with both tensile and compressive forces.

\subsubsection*{External loads and reaction forces of the form}
\begin{figure}
    \centering
    \includegraphics[width = 0.9\columnwidth]{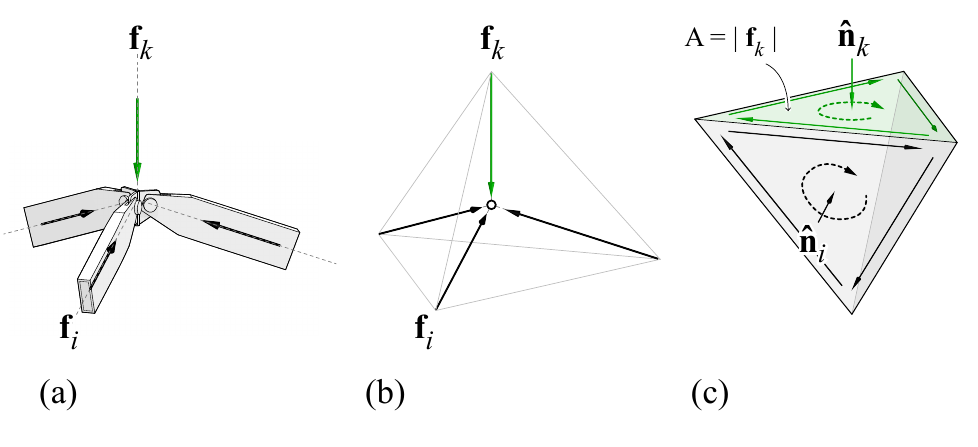}
    \caption{(a) A 3D structural joint with an applied force and internal forces in its members; (b) the form diagram/bar-node representation of the same joint in the context of 3DGS; and (c) the force diagram/polyhedron representing the equilibrium of the same node in 3DGS.}
    \label{fig:01}
\end{figure}

To explain the external loads and reaction forces in the form diagram, consider the example of a polyhedral joint with an externally-applied force $\textbf{f}_k$ of Figure \ref{fig:01}a. This joint can be represented as a group of polyhedral cells in the context of 3DGS as shown in Figure \ref{fig:02}a. Figure \ref{fig:02}a includes four open cells and no closed cell where the open cells represent the applied loads, the reaction forces, and the location of the supports. 

Generally, a group of polyhedral cells with no open cell may represent a self-stressed system of forces with no externally--applied loads \cite{maxwell1864}. Replacing the dashed edges in the form diagram of Figure \ref{fig:01}b with additional members will turn the form into a self-stressed system \cite{CALLADINE1978161}. Since in graphic statics we design the form diagram for externally--applied loads and boundary conditions, so we allow the form diagram to include open polyhedral cells \cite{Akbarzadeh2015}. Subtracting a cell from a group of closed polyhedral cells results in both open and closed cells. We denote the subtracted cell the self-stress polyhedron (SSP) since the group of polyhedrons could be self-stressed otherwise.  

In describing a form diagram, any cell, internal or external, can serve as the SSP. Subtracting the faces of SSP from the group of polyhedrons will leave the adjacent cells open and the rest of the cells closed. The edges connected to the vertices of the chosen SSP represent the vectors of the external loads and the reaction forces. The start point or the end point of the vectors can represent the location of the supports (up to translation). If the SSP is the external polyhedron, all the internal edges connected to the vertices of the external polyhedron will represent the applied loads and reaction forces (Figure \ref{fig:01}b).

\subsection*{The direction of the cells in the form and force diagrams}
\label{sec:dir_cell}
Each cell, in both form and force diagrams, has a direction either towards inside or outside of the cell that is defined by choosing the direction for the SSP/GFP. The direction of the SSP/GFP are either towards inside or outside of the cell. The faces of the cells adjacent to the SSP/GFP will have the same direction as the faces of SSP/GFP. Every other cell in the group, if not adjacent to SSP/GFP, has an opposite direction of its adjacent cell. 

For instance, consider the force diagram of Figure \ref{fig:det_gfp}a; the direction of the GFP is determined by the direction of the externally applied loads and the reaction forces at the supports. The direction of the NFPs will be determined by the direction of GFP; each face of NFP that is shared with the GFP will have the same orientation of the GFP face whereas, the face shared by another NFP will always have an opposite normal direction (Figure \ref{fig:det_gfp}b). 

\subsection{Topological and geometrical properties of the reciprocal polyhedrons}
\label{sec:topo_geo}
\begin{figure}
    \centering
    \includegraphics[width = 0.9\columnwidth]{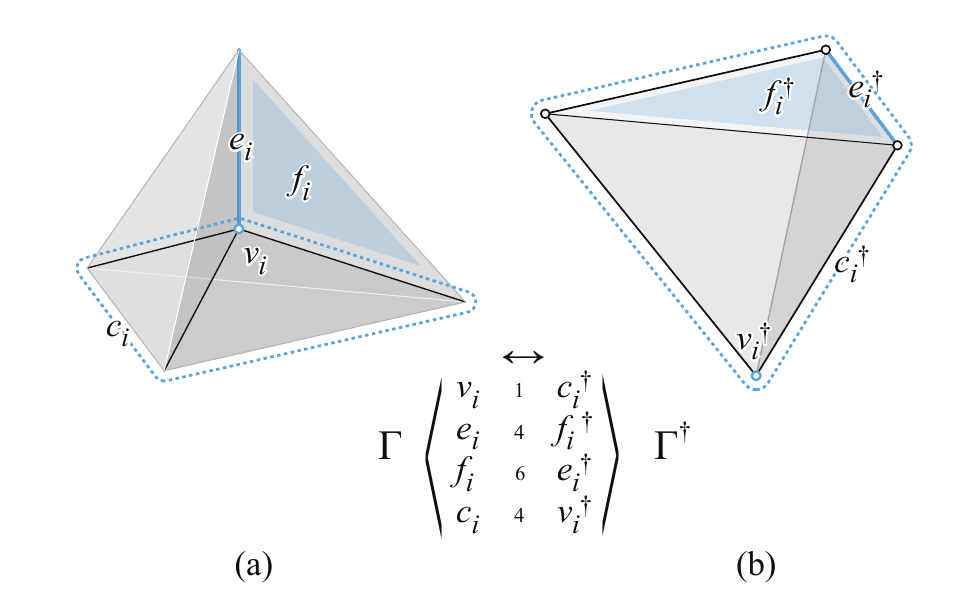}
    \caption{(a) The \textit{primal} diagram $\Gamma$ and (b) its reciprocal diagram $\Gamma^\dagger$ as called \textit{dual} and their corresponding components.}
    \label{fig:02}
\end{figure}

We can use the example of Figure \ref{fig:02} to explain the topological relationship between reciprocal polyhedral diagrams. Let us call the starting diagram the \textit{primal}, $\Gamma$, and the reciprocal polyhedron the \textit{dual}, $\Gamma^\dagger$ (Fig. \ref{fig:02}a, b). The vertices, edges, faces, and cells of the primal are denoted by $v$, $e$, $f$, and $c$ respectively, and the ones of the dual are super-scripted with a dagger ($\dagger$) symbol.

These two diagrams are topologically dual: i.e. the vertices $v$, edges $e$, faces $f$ and cells $c$ of the primal topologically map to the cells $c^\dagger$, faces $f^\dagger$, edges $e^\dagger$ and vertices $v^\dagger$, respectively of the dual diagram \cite{Akbarzadeh2015}. Therefore, the number of the dual elements in both diagrams are the same. For instance, the number of vertices $v$ of the primal is equal to the number of cells $c^\dagger$ in the dual, etc. Moreover, each edge $e$ of the primal is perpendicular to its corresponding face $f^\dagger$ in the dual.

\subsection{Connectivity of the components}
\label{sec:connect}
To formulate the algebraic relationship between the primal and the dual diagrams, the relation between the components of each diagrams should be described algebraically by multiple connectivity matrices for the vertices, edges, faces, and cells of the diagrams.

\begin{figure}[h]
\begin{minipage}[b]{\columnwidth}
    \centering
    \includegraphics[width=0.6\columnwidth]{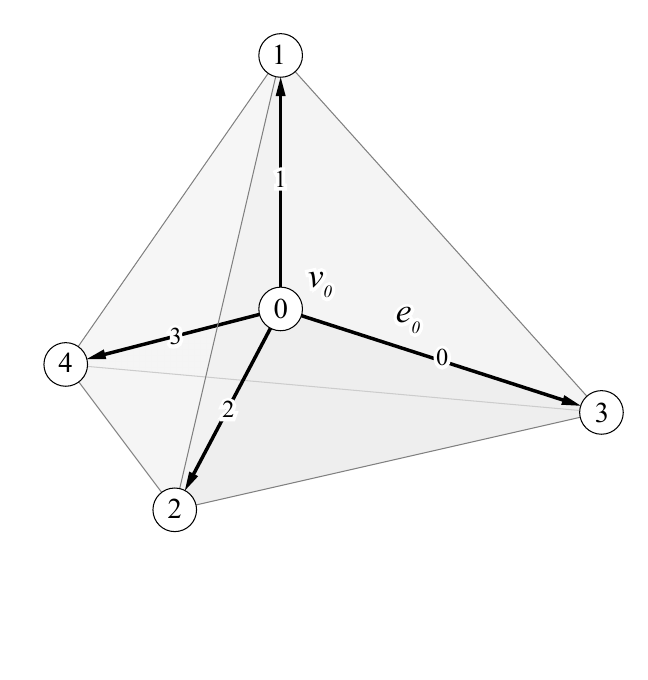}
    \label{fig:image}
  \end{minipage}
  \hfill
  \begin{varwidth}{0.6\columnwidth}
    \centering
    $\textbf{C}_{e\times v} =  
               \bordermatrix{~ & 0 & 1 & 2 & 3 &4\cr
                    0& -1&  0&  0&  1& 0\cr 
                    1& -1&  1&  0&  0& 0\cr 
                    2& -1&  0&  1&  0& 0\cr 
                    3& -1&  0&  0&  0& 1\cr}
            $
    \caption{The primal diagram and the connectivity matrix given by its edges and vertices.}
    \label{fig:e_v}
  \end{varwidth}%
\end{figure}

\subsubsection*{Edge--vertex}
\label{sec:evc}
Let us consider the primal and the dual diagrams of Figure \ref{fig:02}a, b: the primal diagram includes arbitrarily--indexed vertices and the edges pointing from a vertex with a smaller number to a vertex with a bigger number (Figure \ref{fig:e_v}). For the primal diagram, the connectivity matrix between the vertices and edges is a $[e \times v]$ matrix that is shown by $\textbf{C}_{e\times v}$, described as

\[C_{e_{i},v_{j}}=\begin{cases}
+1 & \mbox{if vertex } v_j \mbox{ is the head of edge } e_i\\
-1 & \mbox{if vertex } v_j\mbox{ is the tail of edge } e_i\\0 & \mbox{otherwise.}\\
\end{cases}
\]

Since the edges and vertices of the primal map to faces and cells of the dual, matrix  $\textbf{C}_{e\times v}$ is equal to $\textbf{C}_{{f^\dagger}\times {c^\dagger}}$ that represents the connectivity of the faces and cells of the dual.

\begin{figure}[h]
\begin{minipage}[b]{\columnwidth}
    \centering
    \includegraphics[width=0.6\columnwidth]{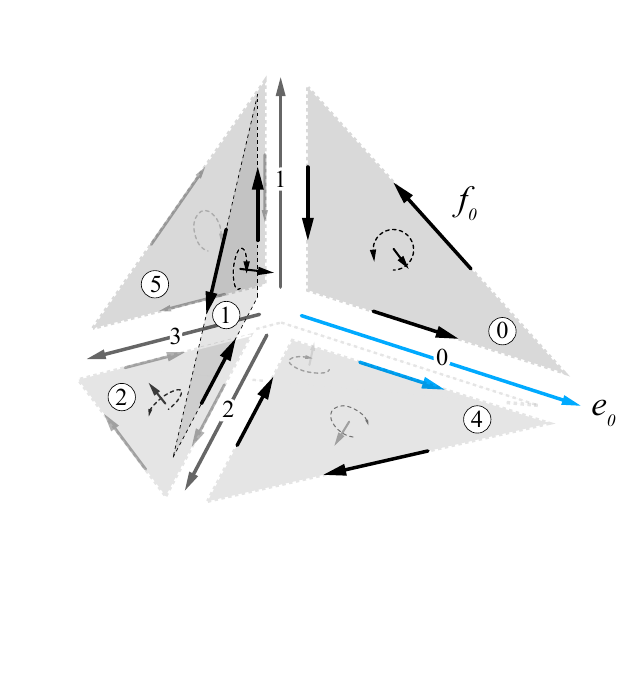}
    \label{fig:e_f1}
  \end{minipage}
  \hfill
  \begin{varwidth}{1\columnwidth}
    \centering
    $\textbf{C}_{e\times f}=
                \bordermatrix{~ & 0 & 1 & 2 & 3 & 4 & 5\cr
                    0&  1&  0&  0&  1&  1&  0\cr
                    1& -1&  1&  0&  0&  0&  1\cr
                    2&  0& -1& -1&  0& -1&  0\cr
                    3&  0&  0&  1& -1&  0& -1\cr}
                $
    \caption{The connectivity of the faces and edges in the primal and its related matrix.}
    \label{fig:e_f}
  \end{varwidth}%
\end{figure}

\subsubsection*{Edge--face}

The connectivity between edges and vertices, $\textbf{C}_{e \times v}$, does not describe the topology of the primal completely, and further connectivity matrices are required to describe the topological relationships among other components of both primal and dual diagrams. Each face $f_i$ of the primal has a unit normal vector $\mathrm{\hat{\textbf{n}}}_i$ where the direction of the normal may be chosen arbitrarily (Figure \ref{fig:e_f}). This direction defines the orientation of the edges $e_i$ on that face using the right-hand rule. 

Therefore, for each edge $e_i$ on the face $f_i$ there are two directions: one given by matrix $\textbf{C}_{e \times v}$  and one defined by the direction of the unit normal of the face $\mathrm{\hat{\textbf{n}}}_i$ (Figure \ref{fig:e_f}). Thus, for the edges and their connected faces in the primal complex, the edge-face connectivity matrix $\textbf{C}_{e \times f}$ is a $[e\times f]$ matrix defined as 

\[C_{e_i, f_j}=\begin{cases}
+1 & \mbox{if edge }e_i\mbox { is an edge of face }f_j \\
-1 & \mbox{if opposite of edge }e_i\mbox { is an edge of face }f_j \\0 & \mbox{otherwise.}\\

\end{cases}\]

Note that matrix $\textbf{C}_{e \times f}$ can also describe the connectivity between the faces $f^\dagger$  and edges $e^\dagger$ of the dual complex and thus equals matrix $\textbf{C}_{{f^\dagger} \times {e^\dagger}}$.



\begin{figure}[h]
\begin{minipage}[b]{\columnwidth}
    \centering
    \includegraphics[width=0.6\columnwidth]{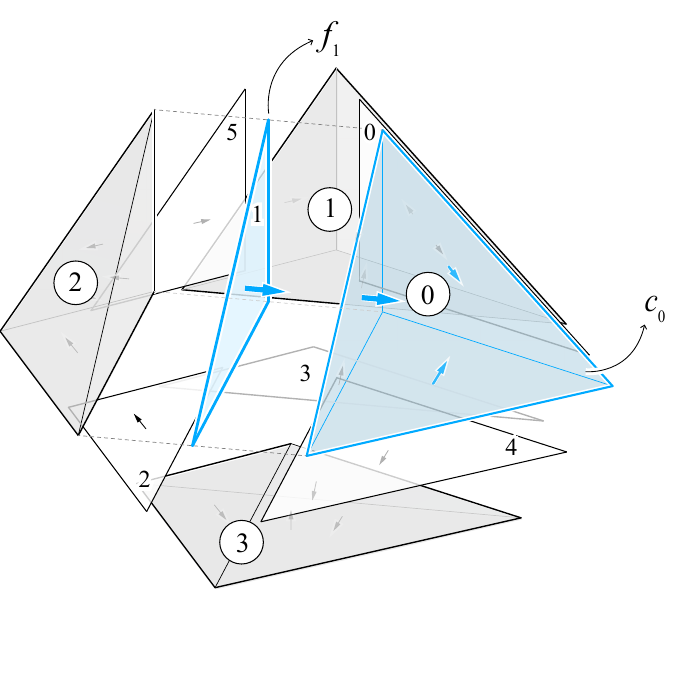}
    \label{fig:f_c1}
  \end{minipage}
  \hfill
  \begin{varwidth}{0.6\columnwidth}
    \centering
    $\textbf{C}_{f\times c} = 
               \bordermatrix{~ & 0 & 1 & 2 & 3\cr
                    0&  1& -1&  0&   0\cr 
                    1&  1&  0& -1&   0\cr 
                    2&  0&  0&  1&  -1\cr
                    3&  0&  1&  0&  -1\cr
                    4& -1&  0&  0&   1\cr
                    5&  0& -1&  1&   0\cr}
            $
    \caption{The connectivity of faces and cells of the primal and its incidence matrix.}
    \label{fig:f_c}
  \end{varwidth}%
\end{figure}

\subsubsection*{Face--cell}
To complete the topological definition of the primal complex, the connectivity between the faces and cells of the primal should be described by an incidence matrix $\textbf{C}_{f \times c}$. 
The direction of each face $f_i$ in the primal was chosen arbitrarily.However, the direction of the cells are predefined as discussed in Section \ref{sec:dir_cell}. We check the direction of face $f_i$ with the direction of the cell $c_j$. Hence, the incidence matrix for the faces and cells can be defined as (Figure \ref{fig:f_c}a, b):  

\[C_{f_i, c_j}=\begin{cases}
+1 & \mbox{if face }f_i\mbox { has the same direction as of }c_j  \\
-1 & \mbox{if face }f_i\mbox { has the opposite direction as of }c_j \\
0 & \mbox{otherwise.}\\

\end{cases}\]

\begin{figure}
    \centering
    \includegraphics[width = .6\columnwidth]{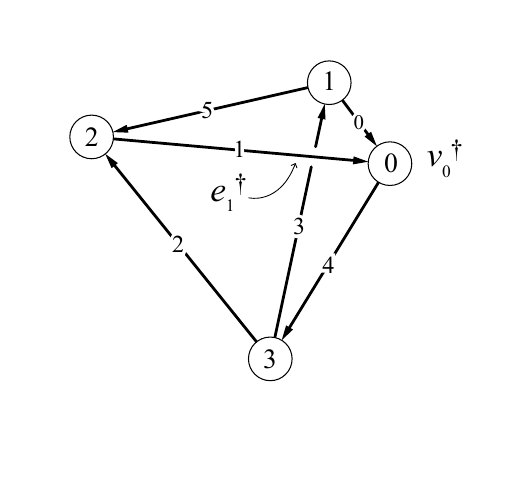}
    \caption{The edge-vertex connectivity of the dual diagram is the same as the face-cell connectivity of the the primal of Figure \ref{fig:f_c}.}
    \label{fig:e_v_dual}
\end{figure}

Since faces $f$ and cells $c$ of the primal correspond to $e^\dagger$ and $v^\dagger$ of the dual, the matrix $\textbf{C}_{f \times c}$ can describe the topological relationship between the unknown edges and vertices of the dual complex and therefore is equal to a $[e^\dagger\times v^\dagger]$ matrix $\textbf{C}_{e^\dagger \times v^\dagger}$ (Figure \ref{fig:e_v_dual}). Note that the direction of the edges of the dual is a result of the face--cell connectivity matrix. For instance, the edges are not necessarily directed from smaller indices to bigger indices.

\subsection{Algebraic reciprocal constraints}
\label{sec:algreccon}
In this section, we describe the algebraic constraints for constructing the dual from the primal complex.  The coordinate difference vectors, $\textbf{u}$, $\textbf{v}$, $\textbf{w}$ can describe the edges of the primal as

\begin{equation}
    \textbf{u}=\textbf{C}_{e\times v}\textbf{x}\quad \textbf{v}=\textbf{C}_{e\times v}\textbf{y}\quad \textbf{w}=\textbf{C}_{e\times v}\textbf{z}
\end{equation}

where $\textbf{x}$, $\textbf{y}$ and $\textbf{z}$ vectors are the $x$-, $y$- and $z$-coordinates of the vertices, and  $\textbf{C}_{e\times v}$ is the incidence matrix for the edges and vertices of the primal. Similarly, $\textbf{u}^\dagger$, $\textbf{v}^\dagger$ and $\textbf{w}^\dagger$ are the coordinate difference vectors that can describe the edges of the dual as

\[\textbf{u}^\dagger=\textbf{C}_{e^\dagger \times v^\dagger}\textbf{x}^\dagger\quad \textbf{v}^\dagger=\textbf{C}_{e^\dagger \times v^\dagger}\textbf{y}^\dagger\quad \textbf{w}^\dagger=\textbf{C}_{e^\dagger \times v^\dagger}\textbf{z}^\dagger.\]

Since vertices and edges of the dual correspond to the cells and faces of the primal, we can write:

\begin{equation}
\label{eq:00}
 \textbf{u}^\dagger=\textbf{C}_{f \times c}\textbf{x}^\dagger\quad \textbf{v}^\dagger=\textbf{C}_{f \times c}\textbf{y}^\dagger\quad \textbf{w}^\dagger=\textbf{C}_{f \times c}\textbf{z}^\dagger,   
\end{equation}

where $\textbf{x}^\dagger$, $\textbf{y}^\dagger$ and $\textbf{z}^\dagger$ are the vector of $x$-, $y$- and $z$-coordinates of the vertices of the dual, and $\textbf{C}_{f \times c}$ is the connectivity matrix of the face and cells of the primal. 

The first set of constraints is imposed by the faces of the dual: around every face ${f_i}^\dagger$, the sum of the coordinate differences of the edges ${\textbf{u}}^\dagger, {\textbf{v}}^\dagger$, and ${\textbf{w}}^\dagger$ has to be zero. The faces ${f_i}^\dagger$ of the dual correspond to edges ${e_i}$ of the primal, and edges ${e_i}^\dagger$ of the dual correspond to the faces ${f_i}$ of the primal. Therefore, the first set of constraints can be described as

\begin{equation}
\label{eq:01}
\textbf{C}_{e \times f}\textbf{u}^\dagger=\textbf{0} \quad \textbf{C}_{e \times f}\textbf{v}^\dagger=\textbf{0} \quad \textbf{C}_{e \times f}\textbf{w}^\dagger=\textbf{0}.   
\end{equation}

\begin{figure}
    \centering
    \includegraphics[width = \columnwidth]{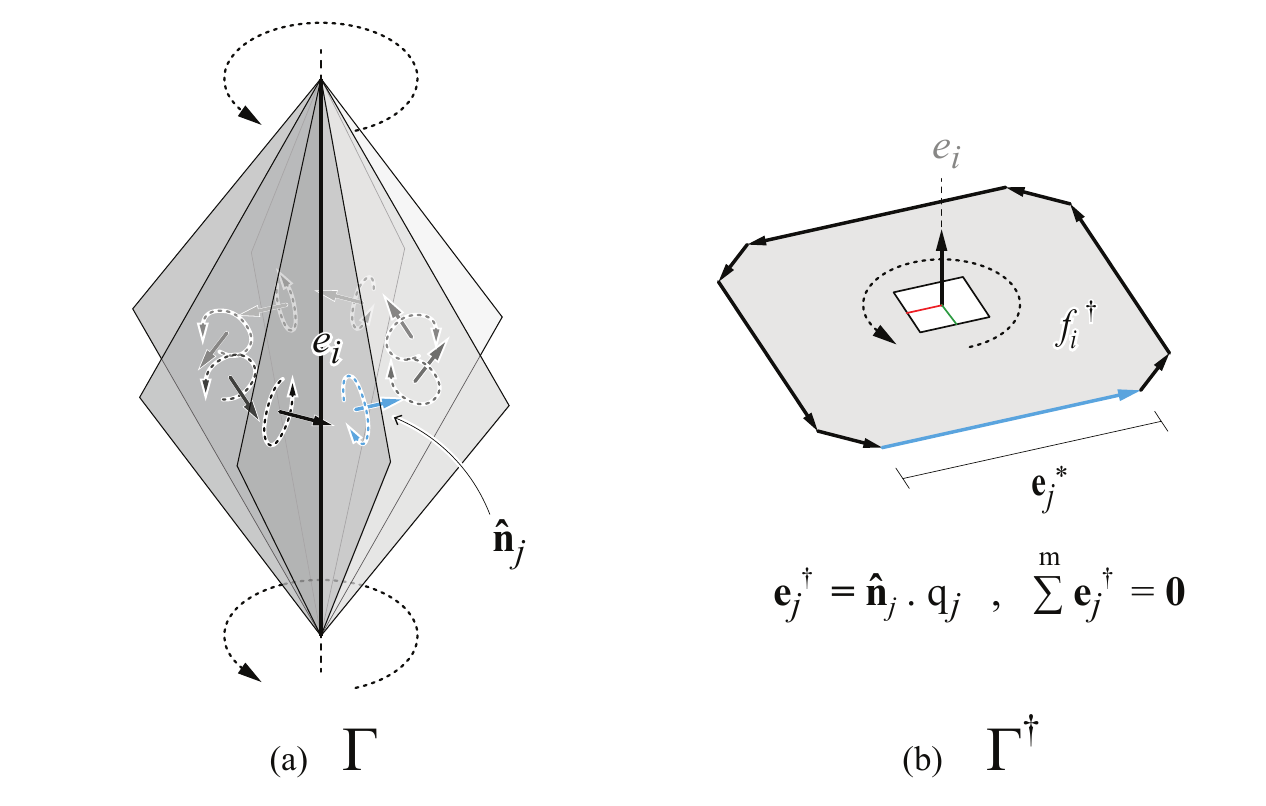}
    \caption{Going around each edge of the primal with its attached faces (a) provides the direction of the edge vectors of the corresponding face (b) in the reciprocal diagram where the sum of the edge vectors must be zero.}
    \label{fig:around_e}
\end{figure}

Moreover, the edges of the dual ${e_i}^\dagger$ and the corresponding normal vectors of the faces of the primal $\mathrm{\hat{\textbf{n}}}_i$ are parallel that establishes the second set of constraints. In other words, around each internal edge $e_i$ and its adjacent faces $f_{i-k}$ in the primal, the sum of the normal vector of the faces $\mathrm{\hat{\textbf{n}}}_i$ multiplied by the length $q_i$ of the corresponding edge ${e}^\dagger_i$ in the dual diagram should be the zero vector (Figure \ref{fig:around_e}).

Let $\textbf{N}_x$, $\textbf{N}_y$ and $\textbf{N}_z$ be the $[f\times f]$ diagonal matrices whose diagonal entries are the $x$-, $y$- and $z$-coordinates (respectively) of the chosen unit normal vectors of the faces of the primal. Further, let $\textbf{q}$ denote the \textit{scale vectors} or the \textit{force densities} that define the lengths of the edges of the dual \cite{Schek1974115}. Therefore, the second set of constraints can be written as 
\begin{equation}
\label{eq:02}
    \textbf{u}^\dagger=\textbf{N}_x\textbf{q}\quad \textbf{v}^\dagger=\textbf{N}_y\textbf{q}\quad\textbf{w}^\dagger=\textbf{N}_z\textbf{q}.
\end{equation}

Combining Eq. \ref{eq:01} and Eq. \ref{eq:02} results in
\begin{equation}
\label{eq:03}
\textbf{C}_{e \times f}\textbf{N}_x\textbf{q}=\textbf{0}\quad \textbf{C}_{e \times f} \textbf{N}_y\textbf{q}=\textbf{0} \quad \textbf{C}_{e \times f} \textbf{N}_z\textbf{q}=\textbf{0}.
\end{equation}

We call the $[3e\times f]$ matrix
\[\textbf{A}=\left(\begin{array}{c}
\textbf{C}_{e \times f} \textbf{N}_x\\ \hline \textbf{C}_{e \times f}\textbf{N}_y \\ \hline \textbf{C}_{e \times f} \textbf{N}_z \end{array}\right)\]
the \textit{equilibrium} matrix of the problem. Any solution of the equation system

\begin{equation}
\label{eq:04}
 \textbf{A}\textbf{q}=\textbf{0}   
\end{equation}

gives us a possible vector (force density) for the edge of the dual, and hence a possible solution to the problem of constructing the dual.

\subsection{Solutions}
\label{sec:solutions}
The dimension of the solutions $\textbf{q}$ satisfying Eq. \ref{eq:04} is equal to the dimension of the right nullspace of $\textbf{A}$. I.e. if we have $r$ independent equations, the dimension of the right null space is equal to $f-r$. The $r$ is the number of independent equations of \ref{eq:04} which is equal to the rank of the equilibrium matrix $\textbf{A}$. For instance, Figure \ref{fig:02}a shows a primal polyhedral system that includes tetrahedral cells with the SSP chosen as the exterior cell with equilateral triangle faces. The matrix $\textbf{A}$ for this primal will be as follows:

\[\textbf{A}=\left(\begin{array}{cccccc}
-\frac{1}{2\sqrt{3}} & \frac{\sqrt{3}}{2} & -\frac{1}{\sqrt{3}} & 0 & 0 & 0\\
0 & 0 & -\frac{\sqrt{3}}{2} & 0 & 0 & \frac{\sqrt{3}}{2}\\
0 & 0 & \frac{1}{\sqrt{3}} & 0 & \frac{1}{2\sqrt{3}} & -\frac{\sqrt{3}}{2}\\
\frac{1}{2\sqrt{3}} & 0 & 0 & 0 & -\frac{1}{2\sqrt{3}} & 0\\
-\frac{1}{2} & \frac{1}{2} & 0 & 0 & 0 & 0\\
0 & -\frac{1}{2} & 0 & 1 & 0 & -\frac{1}{2}\\
0 & 0 & 0 & 0 & -\frac{1}{2} & \frac{1}{2}\\
\frac{1}{2} & 0 & 0 & -1 & \frac{1}{2} & 0\\
\frac{\sqrt{2}}{\sqrt{3}} & 0 & -\frac{\sqrt{2}}{\sqrt{3}} & 0 & 0 & 0\\
0 & 0 & 0 & 0 & 0 & 0\\
0 & 0 & \frac{\sqrt{2}}{\sqrt{3}} & 0 & -\frac{\sqrt{2}}{\sqrt{3}} & 0\\
-\frac{\sqrt{2}}{\sqrt{3}} & 0 & 0 & 0 & \frac{\sqrt{2}}{\sqrt{3}} & 0
\end{array}\right)\]

Matrix $\textbf{A}$ of this example is a $12 \times 6$ matrix of rank $5$ ($r = 5$). The dimension of the right nullspace, $(f-r)$, is $6-5$ which equals to $1$ and therefore, there is a unique solution (up to scaling and translation) (Figure \ref{fig:02}b). 

\subsection{Constructing the geometry of the dual}
\label{sec:condual}
We developed two approaches to construct the geometry of the dual, and we will explain both methods in this section. The first approach is purely algebraic, whereas the second approach involves a graph--search algorithm to construct the geometry of the dual.  

\subsubsection*{Algebraic approach}

Once we find a solution $\textbf{q}$ for Eq. \ref{eq:04}, we can compute the coordinate difference vectors $\textbf{u}^\dagger$, $\textbf{v}^\dagger$, and $\textbf{w}^\dagger$ using Eq. \ref{eq:02}. 

In order to construct the geometry of the dual, we need to compute the coordinates $\textbf{x}^\dagger$, $\textbf{y}^\dagger$, $\textbf{z}^\dagger$ of the vertices of the dual. Given a solution $\textbf{q}$, the vectors $\textbf{u}^\dagger$, $\textbf{v}^\dagger$ and $\textbf{w}^\dagger$ are determined, and from \ref{eq:01} and \ref{eq:02} we have

\begin{equation}
    \textbf{N}_x\textbf{q}=\textbf{C}_{f \times c}\textbf{x}^\dagger\quad \textbf{N}_y\textbf{q}=\textbf{C}_{f \times c}\textbf{y}^\dagger\quad \textbf{N}_z\textbf{q}=\textbf{C}_{f \times c}\textbf{z}^\dagger.
\end{equation}
 Multiplying both sides with the transpose $\textbf{C}_{c\times f}$ of the incidence matrix $\textbf{C}_{f \times c}$ results in the Laplacian matrix $\textbf{L}^\dagger$ on the right side

\[
\textbf{L}^\dagger = \textbf{C}_{c \times f} \textbf{C}_{f \times c},
\]

and therefore, 
\begin{equation}
\begin{split}
    \textbf{C}_{c \times f}\textbf{N}_x\textbf{q}=\textbf{L}^{\dagger}\textbf{x}^\dagger, \textbf{C}_{c \times f}\textbf{N}_y\textbf{q}=\textbf{L}^{\dagger}\textbf{y}^\dagger,\\
     \textbf{C}_{c \times f}\textbf{N}_z\textbf{q}=\textbf{L}^{\dagger}\textbf{z}^\dagger.
     \end{split}
\end{equation}

The $[c\times c]$ Laplacian matrix $\textbf{L}^\dagger$ is a positive semi-definite matrix, and it is not invertible. In fact, the translation vectors $\textbf{u}^\dagger$, $\textbf{v}^\dagger$, and $\textbf{w}^\dagger$ need a chosen point in the three-dimensional space to result in a unique solution for $\textbf{x}^\dagger$, $\textbf{y}^\dagger$ and $\textbf{z}^\dagger$. Therefore, we start by choosing a vertex $v_0^\dagger$ of the dual as the starting point of the construction and set its coordinates to be all zeros ($0$). Once these coordinates are set, the vectors $\textbf{u}^\dagger$, $\textbf{v}^\dagger$, and $\textbf{w}^\dagger$ uniquely determine $\textbf{x}^\dagger$, $\textbf{y}^\dagger$ and $\textbf{z}^\dagger$ and the whole geometry of the dual.




We formulate the above discussion algebraically as follows: consider the $[1\times c]$ row vector $\boldsymbol{\sigma}$ whose first entry is 1, and the rest of the entries are all 0. Then, the solutions of the linear equations

\begin{equation*}
    \boldsymbol{\sigma}\cdot \textbf{x}^\dagger=0\quad\boldsymbol{\sigma}\cdot \textbf{y}^\dagger=0\quad\boldsymbol{\sigma}\cdot \textbf{z}^\dagger=0
\end{equation*}

are exactly those $\textbf{x}^\dagger$, $\textbf{y}^\dagger$, and $\textbf{z}^\dagger$ vectors whose first entry is zero ($0$). We add this linear equation to the Eq. \ref{eq:00}, obtaining a $[(f+1) \times c]$ matrix \[\textbf{C}^{\boldsymbol{\sigma}}_{f\times c}=\left(\begin{array}{c} \boldsymbol{\sigma}\\ \hline \textbf{C}_{f\times c}\end{array}\right)\] and a $[(f+1) \times 1]$ column vectors

\begin{equation*}
    \textbf{u}_{\boldsymbol{\sigma}}^\dagger=\left(\begin{array}{c}0\\ \hline \textbf{u}^\dagger\end{array}\right)\quad \textbf{v}_{\boldsymbol{\sigma}}^\dagger=\left(\begin{array}{c}0\\ \hline \textbf{v}^\dagger\end{array}\right)\quad\textbf{w}_{\boldsymbol{\sigma}}^\dagger=\left(\begin{array}{c}0\\ \hline \textbf{w}^\dagger\end{array}\right).
\end{equation*}


The solutions of the equation systems

\begin{equation}
\label{eq:05}
\textbf{C}^{\boldsymbol{\sigma}}_{f\times c}\textbf{x}^\dagger=\textbf{u}_{\boldsymbol{\sigma}}^\dagger\quad\textbf{C}^{\boldsymbol{\sigma}}_{f\times c}\textbf{y}^\dagger=\textbf{v}_{\boldsymbol{\sigma}}^\dagger\quad\textbf{C}^{\boldsymbol{\sigma}}_{f\times c}\textbf{z}^\dagger=\textbf{w}_{\boldsymbol{\sigma}}^\dagger
\end{equation}

are exactly those solutions of the original Eqs. \ref{eq:00} whose first coordinates are zero ($0$). The columns of the matrix $\textbf{C}^{\boldsymbol{\sigma}}_{f\times c}$ are linearly independent, hence the equation systems of \ref{eq:05} has a unique solution. This unique solution can be computed by using the Moore-Penrose inverse of the matrix $\textbf{C}^{\boldsymbol{\sigma}}_{f\times c}$ that is denoted by $\textbf{M}^{\boldsymbol{\sigma}}_{{f\times c}}$ as
\[\textbf{M}^{\boldsymbol{\sigma}}_{f\times c}=\left(\textbf{C}^{\boldsymbol{\sigma}}_{c\times f}\textbf{C}^{\boldsymbol{\sigma}}_{f\times c}\right)^{-1}\textbf{C}^{\boldsymbol{\sigma}}_{c\times f}.\]
Here, $\textbf{C}^{\boldsymbol{\sigma}}_{c\times f}$ denotes the transpose of the matrix $\textbf{C}^{\boldsymbol{\sigma}}_{f\times c}$.

Explicitly, the unique solutions are given as 

\begin{equation*}
    \textbf{x}^\dagger=\textbf{M}^{\boldsymbol{\sigma}}_{f\times c}\cdot \textbf{u}_{\boldsymbol{\sigma}}^\dagger\quad
    \textbf{y}^\dagger=\textbf{M}^{\boldsymbol{\sigma}}_{f\times c}\cdot \textbf{v}_{\boldsymbol{\sigma}}^\dagger\quad
    \textbf{z}^\dagger=\textbf{M}^{\boldsymbol{\sigma}}_{f\times c}\cdot \textbf{w}_{\boldsymbol{\sigma}}^\dagger
\end{equation*}


Note that the square matrix $\textbf{C}^{\boldsymbol{\sigma}}_{c\times f}\textbf{C}^{\boldsymbol{\sigma}}_{f\times c}$ is very similar to the Laplacian of the original incidence matrix $\textbf{C}_{f\times c}$ in that all the entries but the top left are equal. We remark that the square matrix $\textbf{C}^{\boldsymbol{\sigma}}_{c\times f}\textbf{C}^{\boldsymbol{\sigma}}_{f\times c}$ is a positive definite matrix.

\subsubsection*{Graph--search approach}
We can also avoid the algebraic approach in constructing the dual to find the tree graph of the dual using the face-cell topology of the primal. The tree graph includes paths from a chosen vertex to all other vertices with no closed loops that can be found using Breadth-First-Search (BFS) algorithm. 

To construct the geometry, we can assign particular $x-$, $y-$, $z-$ coordinates to a vertex of the dual $v_0^\dagger$ and use it as the starting point of the construction. This step is the same as choosing a start point in the previous section. Then, we find all paths including segments of the dual parsed from vertex $v_0^\dagger$ to reach to each vertex $v_i^\dagger$. Each segment in each path includes a start and end vertex corresponding to two cells with a shared face $f_i$ in the primal. Each segment must be weighted by the force density $q_i$, and it has the direction of the normal $\mathrm{\hat{\textbf{n}}}_i$ of the corresponding face in the cell reciprocal to the end vertex. 

\begin{figure}
    \centering
    \includegraphics[width = 1.\columnwidth]{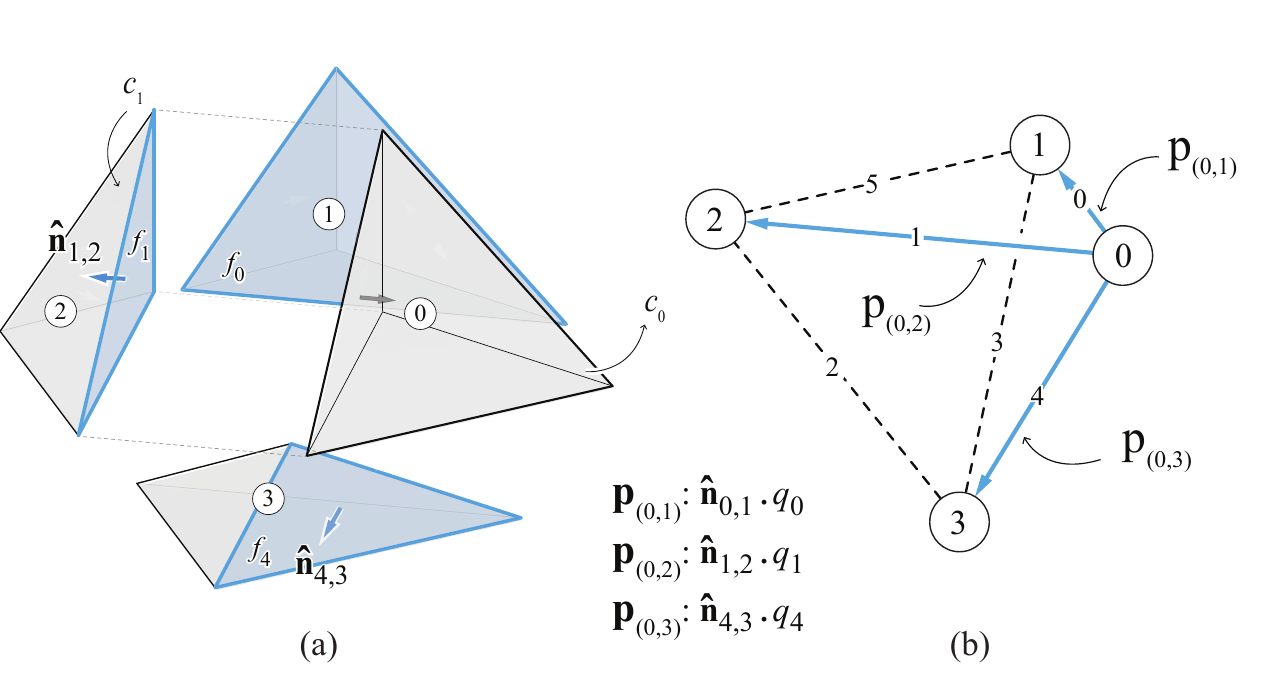}
    \caption{Graph--search approach to construct the geometry of the dual: (a)}
    \label{fig:05}
\end{figure}

For instance, Figure \ref{fig:05}b shows three paths to find all the coordinates of the vertices of the dual for the primal of Figure \ref{fig:02}a. The path $p_{(0,1)}$, in this case, includes one segment where the length $q_0$ is multiplied by the direction of the normal of the face $f_0$ in the cell $c_1$.       

\subsection{Primal as the force diagram}
\label{sec:prime_force}

The previous sections described an algebraic approach to construct the reciprocal diagram for a given primal. This method is a bi-directional approach in the context of 3D graphic statics. I.e. the primal can be considered as either the form or the force diagram. If the primal is considered as the force diagram, the GFP should be defined to find the direction of the cells for the whole complex. The dual will be the form of a structure where the configuration of internal and external forces are in equilibrium according to the primal.

\begin{figure}[h]
    \centering
    \includegraphics[width = 1.\columnwidth]{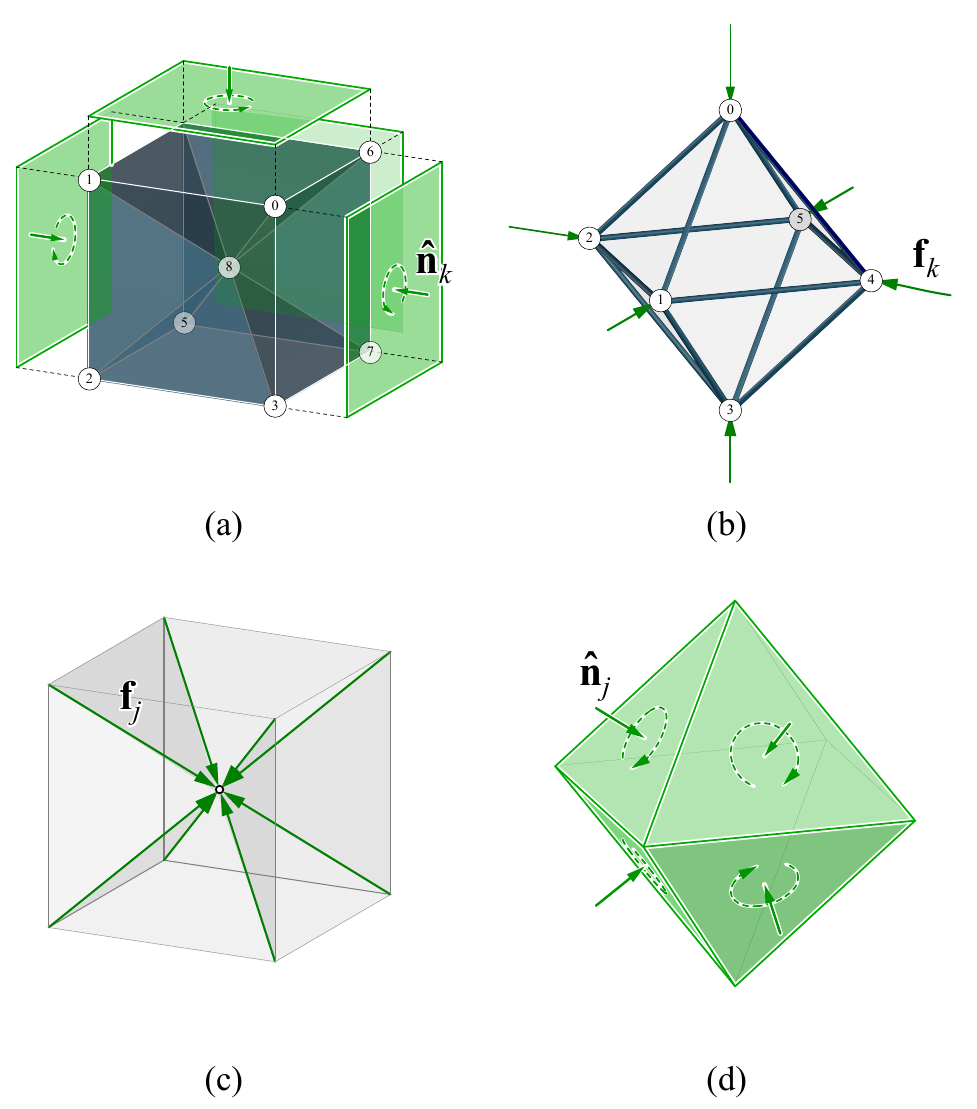}
    \caption{(a) A group of polyhedral cells as the primal where GFP is the external cell; (b) the dual complex representing the form diagram resulted from algebraic approach; (c) the same polyhedrons as the form diagram where the vertices of the external polyhedron defines the externally--applied loads; and (d) its reciprocal force diagram.}
    \label{fig:iass_01}
\end{figure}

For instance, Figure \ref{fig:iass_01}a illustrates a group of closed polyhedral cells representing the force diagram as the primal. The GFP is the external force polyhedron with face normals pointing toward inside of the cell. All other NFPs are convex, and their direction can be defined by the GPF. The algebraic formulation finds the dual as a  compression/tension-only structural form illustrated in Figure \ref{fig:iass_01},b.

\subsubsection*{Tensile vs compressive members}
\label{sec:tencom}

For a primal as the force diagram the type of internal forces in the members of the dual should be defined. To find the direction of the force, we need to compare the topological and geometric directions of the edge $e_i^\dagger$ of the dual which has vertices $v_j^\dagger$ and $v_k^\dagger$, and the order of the vertices is given by the connectivity matrix $\textbf{C}_{f\times c}$. The geometric direction of the vector $\textbf{e}_{i}^\dagger$ is given by the vector starting from $v_j^\dagger$ and ending at $v_k^\dagger$. The direction of a vector from the topological order is the direction of the normal $\mathrm{\hat{\textbf{n}}}_i$ of the face $f_i$ in the cell $c_k$. Therefore   

\begin{equation}
\psi_{e_i} = \mathrm{\textbf{e}_{i}^\dagger}\cdot\mathrm{\textbf{\^n}_{k}}
\end{equation}

where $\psi$ is the dot product of the two directions. According to the following definition we can find the type of internal force in each member:

\[\mbox{if GFP}\begin{cases}
\mbox{negative (inward)}
    \begin{cases}
    \mbox{if }\psi_{e_i} > 0\mbox{ : }e_i^\dagger\mbox{ is compressive}\\
    \mbox{if }\psi_{e_i} < 0\mbox{ : }e_i^\dagger\mbox{ is tensile}
    \end{cases}\\
\mbox{positive (outward)}
    \begin{cases}
    \mbox{if }\psi_{e_i} > 0\mbox{ : }e_i^\dagger\mbox{ is tensile}\\
    \mbox{if }\psi_{e_i} < 0\mbox{ : }e_i^\dagger\mbox{ is compressive}
\end{cases}
\end{cases}\]

For instance, if the GFP is the external cell, and its direction is inward, then the direction of all the NFPs are consistent and inward. Therefore, the topological direction matches the geometric direction. In such cases, simply the sign of $q$ can define whether the member is in tension or compression. If $q_i$ corresponding to the length of the edge $e_i^\dagger$ is positive, then the edge $e_i^\dagger$ is a compressive member, and if it is negative it will be a tensile member. 

\begin{figure}[h]
    \centering
    \includegraphics[width = .9\columnwidth]{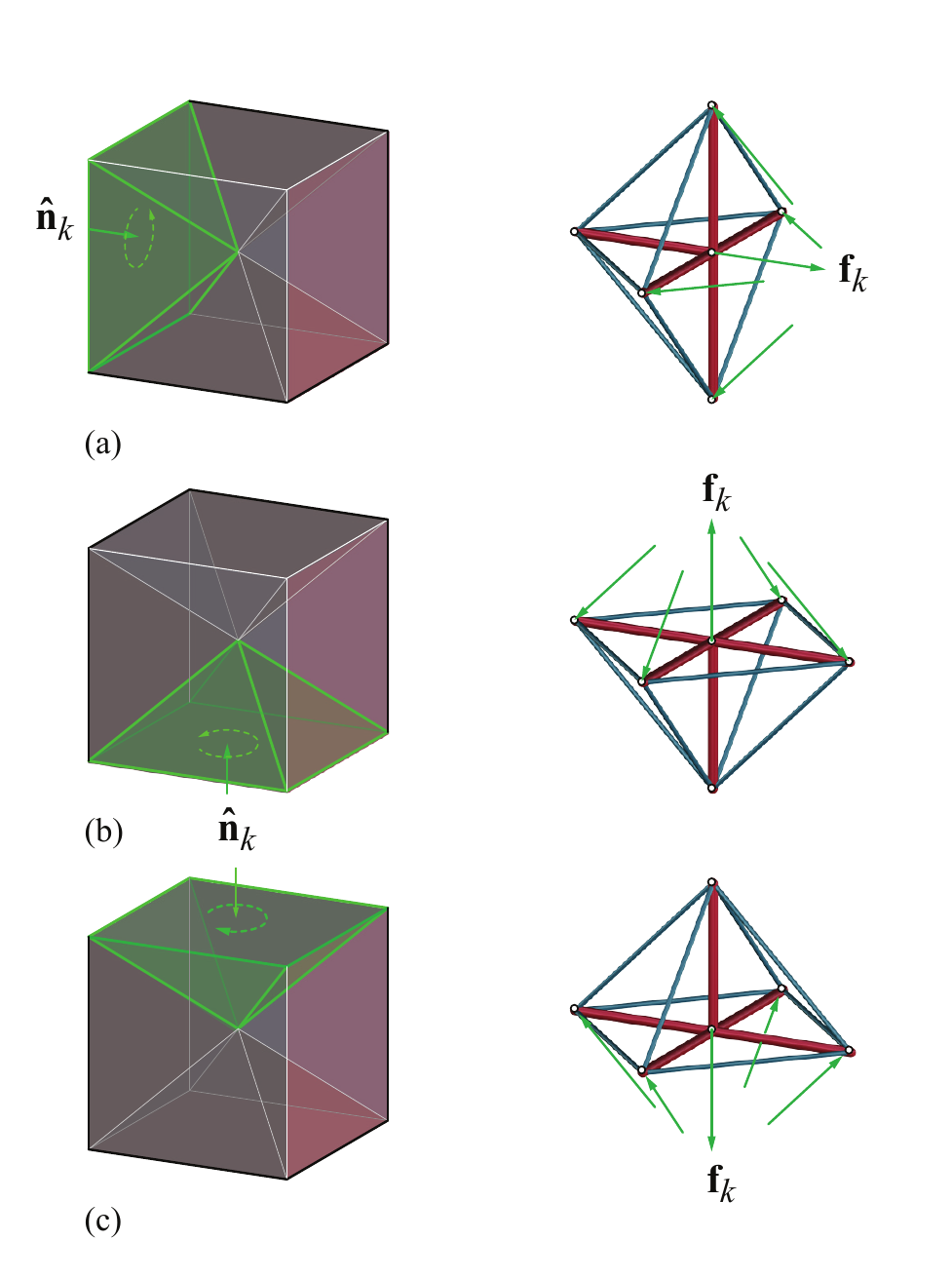}
    \caption{Choosing a different GFP results in compression and tension combined systems.}
    \label{fig:tent-comp}
\end{figure}

Therefore, if all the $q_i$ of a solution vectors $\textbf{q}$ are positive, then the dual is a compression-only system, and if all negative, all the edges are tensile depending on the direction of the GFP (Figure \ref{fig:iass_01}a,b). Choosing any other cell as the GFP results in a form with combined tensile and compressive forces (Fig. \ref{fig:tent-comp}).

\subsection{Primal as the form diagram}
\label{sec:primal_as_form}
The primal can also be considered as the form diagram. In this case, the SSP needs to be chosen to define the external loads and the reaction forces (Figure \ref{fig:iass_01}b). Once the SSP is chosen, the edges connected to the vertices if the SSP represent the applied forces on the form. The same algebraic method can be used to construct the force diagram for a given form; the equilibrium equations will be written around all edges except the edges of SSP. Figure \ref{fig:iass_01}c shows a primal as the form diagram where the SSP is the exterior polyhedron. The resulting diagram of Figure \ref{fig:iass_01}d is the force diagram representing the force magnitudes and the equilibrium of the primal.   

\subsection{The degrees of geometric and static (in)determinacy}
\label{sec:dsi}
If the primal is the form diagram, the dimension of the right nullspace of the equilibrium matrix $\textbf{A}$, $(f-r)$, in fact, is the degree(s) of geometric (in)determinacy of the dual complex that is the force diagram. Note, that the geometric degrees of indeterminacy of the dual is the degrees of static (in)determinacy of the primal complex.

This number is always a non-negative integer: if it is zero ($f-r = 0$), this means that the only solution of Eq. \ref{eq:04} is a zero vector ($\textbf{q}=\textbf{0}$) where the dual collapses into a single point which is not considered as a solution in the context of 3DGS.

If the degree equals one ($f-r = 1$), the set of solutions of Eq. \ref{eq:04} is one-dimensional, that is unique up to scaling. In this case, a non-zero value of a coordinate $q_i$ of the solution $\textbf{q}$ determines the values of the rest of the coordinates. Simply put, there is only one family of solutions for the dual, and therefore, the form is statically determinate. Figure \ref{fig:02}, \ref{fig:iass_01}, and \ref{fig:tent-comp} show examples of input diagrams whose the duals are geometrically determinate. If the primal, is the form diagram, then it is statically determinate. 

If the degree or the dimension of the right nullspace is more than one ($f-r > 1$), there exist at least two solutions up to scaling and translation. That is, the dual is statically indeterminate. 

If the primal complex is the force diagram, then the geometric degrees of (in)determinacy of the dual represents a family of structural forms that are in equilibrium given the primal force distribution. For instance, Figure \ref{fig:dof1} shows an example of an input complex as the force diagram with several significantly different duals/forms, hence the dual is geometrically indeterminate.

%% file: implementation.tex
\begin{figure}
    \centering
    \includegraphics[height = .9\textheight]{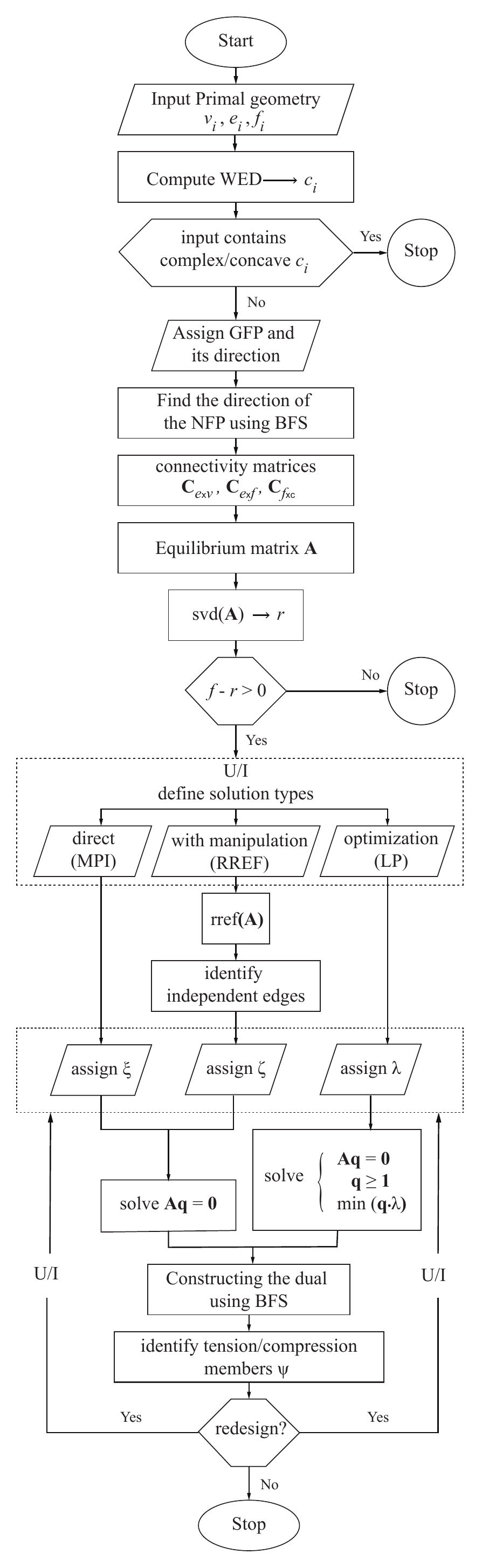}
    \caption{The computational flowchart for algebraic reciprocal construction}
    \label{fig:flowchart}
\end{figure}

\section{Computational setup}
\label{sec:comp_set}
In this section, we explain the computational setup as it is illustrated in the flowchart of Figure \ref{fig:flowchart}. In this flowchart, the primal is the force diagram, and the algebraic method is used for form finding. However, the same setup can be used for structural analysis if the primal is the form diagram as explained in section \ref{sec:primal_as_form}.  

\subsection{Constructing the Winged-Edge data structure}
The computational setup has been implemented in the environment of Rhinoceros software \cite{mcneel2015rhinoceros} and the input includes series of connected planar faces representing a group of polyhedral cells. The first step in the computational setup is to define the topology of the primal complex including the cells, edges and faces and construct their connectivity matrices. Winged-Edge data structure (WED) or alike can be used to find all possible convex cells and the topological relationships  \cite{Akbarzadeh2015, Baumgart1975}. One of the current limitations of this implementation is that the input cannot accept complex (self-intersecting) faces, and therefore, it can only find convex polyhedral cells.  

\subsection{Assigning GFP and its direction}
The method we propose in this paper is applicable to both form finding and analysis. In the form--finding approach, the user should define the GFP to find the direction of the cells in the primal complex. For compression/tension--only form finding, the external polyhedron is chosen as the GFP. The direction of the internal cells are found by the direction of the GFP as explained in Section \ref{sec:dir_cell}. 

\subsection{Solving equilibrium equations}
Writing the equilibrium equations around the edges of the primal (except the edges of the global cell/exterior cell) results in the equilibrium matrix $\textbf{A}$. In the following sections we demonstrate several methods to solve Eq. \ref{eq:04} for $\textbf{q}$ and highlight the advantages of using each method.  

\subsubsection{Moore-Penrose inverse method}
The equilibrium matrix $\textbf{A}$ is usually not invertible. We can use the Moore-Penrose inverse (MPI) of $\textbf{A}$ denoted by $\textbf{A}^{+}$ to solve Eq. \ref{eq:04}. The $\textbf{A}^{+}$ of $\textbf{A}$ satisfies the following matrix equations

\begin{equation*}
\textbf{AA}^{+}\textbf{A}=\textbf{A} \quad\mbox{,}\quad \textbf{A}^{+}\textbf{AA}^{+}=\textbf{A}.
\end{equation*}

From the first equality, any vector $\textbf{q}$ of the form
\begin{equation}
\label{eq:06}
\textbf{q}=(\textbf{I}-\textbf{A}^+ \textbf{A})\xi\
\end{equation}

 solves the linear equation system Eq. \ref{eq:04} where $\textbf{I}$ is the $[f\times f]$ identity matrix and $\xi$ is any $[f\times 1]$ column vector. In fact, all solutions of the Eq. \ref{eq:04} will have the form of Eq. \ref{eq:06} \cite{Moore1920,Penrose1955}. Hence, MPI can generate all the solutions of the equilibrium equations. Note, that the user can choose the components of the vector $\xi$. For instance, assigning 1 to all components gives us a dual solution with a well-distributed edges lengths. Moreover, for primal input with multiple axes of symmetry, this approach results in symmetrical dual solution (Fig. \ref{fig:dof1}a,b). However, the user cannot specify certain edge lengths to particular edges of the dual. In order to address this limitation, we propose the following approach to solve the equilibrium equations.   


\subsubsection{Reduced row echelon form approach}
\label{sec:rref}
Since the dimension of the solutions of the equilibrium equations is $f-r$, therefore, we have exactly $f-r$ independent equations in the equilibrium matrix. This means that we can specify the length of $f-r$ edges of the dual and the rest of the edges will be determined accordingly. Simply put, a user can interact with $f-r$ independent edges to manipulate the geometry of the dual.

The reduced row echelon from (RREF) $\textbf{A}^{rref}$ of the matrix $\textbf{A}$ identifies the independent edges of the dual, because the rank of $\textbf{A}$ equals to the number of pivots in $\textbf{A}^{rref}$. The independent edges correspond to those columns of $\textbf{A}^{rref}$ where there is no pivot. The coordinates corresponding to these columns can be represented by a $[(f-r)\times 1]$ column vector $\zeta$. Any chosen value for the components of the  $\zeta$ will determine the geometry of the dual. 

To address the approach mathematically, we reorder the columns of the $\textbf{A}^{rref}$ matrix so that the pivots are in the main diagonal. Then we exclude all zero rows of the matrix to obtain a $[r\times f]$ matrix $\textbf{A}^{rref}_{r\times f}$. The first $r$ columns of this matrix form the $[r\times r]$ identity matrix, $\textbf{I}$. We denote the $[r\times (f-r)]$ matrix formed by the last $f-r$ columns by $\textbf{B}$, so that

\begin{equation}
\label{eq:070}
    \textbf{A}^{rref}_{r\times f}=\left(\textbf{I}|\textbf{B}\right).
\end{equation}

We can use $\textbf{A}^{rref}_{r\times f}\textbf{q}=\textbf{0}$ as the new equilibrium matrix in Eq. \ref{eq:04} as 

\begin{equation}
\label{eq:07}
    \textbf{A}^{rref}_{r\times f}\textbf{q}=\textbf{0}.
\end{equation}
The solutions of the Eq. \ref{eq:07} are the same as the solutions of Eq. \ref{eq:04}, except that the coordinates of the solution vector are reordered as the last $f-r$ coordinates correspond to the independent edges. 

We denote the $[r \times 1]$ column matrix corresponding to the first $r$  coordinates of $\textbf{q}$ by $\textbf{q}_r$ and the $[(f-r)\times 1]$ column matrix corresponding to the last $f-r$ columns by $\zeta$. Using these notations, the equation system \ref{eq:070} becomes
\[\textbf{I}\textbf{q}_r+\textbf{B}\zeta=\textbf{0}.\]
Therefore, the vector $\zeta$ which corresponds to the length of the independent edges determines the rest of the solution vector:
\[\textbf{q}_r=-\textbf{B}\zeta.\]
As a result the user can choose the length of the independent edges to manipulate the geometry of the dual. 

Although any (positive/negative) values can be chosen for the independent edges, there is no guarantee that if all the independent edges have positive values the rest of the edges will also be positive and the resulting geometry will be a  compression/tension--only system (edges with positive lengths). To address this limitation, we suggest using linear programming approach to solve the equilibrium matrix.   


\subsubsection{Linear programming approach}
We can use the following linear optimization setup to find a dual diagram with all positive edge lengths:

\begin{equation}
    \label{eq:08}
    \mbox{Solve}\quad \begin{cases} \textbf{Aq}=\textbf{0}\\ \textbf{q}\geq \textbf{1}\\ \min(\textbf{q}\cdot \lambda)\end{cases}
\end{equation}

where $\textbf{1}$ is the $[f\times 1]$ vector whose all coordinates are one ($1$) and $\lambda$ is a vector that can be chosen by the user. The solution of this linear programming setup is a solution vector $\textbf{q}$ of the equilibrium equation \ref{eq:04} whose coordinates are at least one ($1$) minimizing the objective function 
\[\textbf{q}\cdot \lambda=\sum_{i=1}^f q_i\lambda_i.\] 
Various linear programming software or packages can be used to solve this linear optimization problem. 


Note that Eq. \ref{eq:04} may not always have a positive solution.  However, if there are positive solutions, then we can find one by using the linear programming approach given that $\lambda>\textbf{0}$.

In addition, different $\lambda$ vectors yield different objective functions. For instance, the objective function given by $\lambda=\textbf{1}$ is the sum of the lengths of the edges of the dual. Hence the solution of Eq. \ref{eq:08} is a solution which minimizes the total edge lengths of the dual given that all edges are of length at least one ($1$). This method can be used to generate structural solutions in 3D with the minimum load-paths that can significantly reduce the use of materials in the structure \cite{Beghini2013b}.   

\subsubsection{Improving the speed and precision}
The speed and precision of the mentioned approaches to solve Eq. \ref{eq:04} can be significantly improved by eliminating redundant rows of $\textbf{A}$ prior to any computation. Authors, in an earlier publication, developed two methods to eliminate redundant rows of $\textbf{A}$ that results in a matrix with only $2(e-v)$ rows, instead of the original $3e$ number of rows \cite{akbarzadeh2018a}. 

 \subsection{Constructing the dual}
Once a solution vector of Eq. \ref{eq:04} is obtained, we can construct the geometry of the dual either using the algebraic method or the graph-search method as described in Section \ref{sec:condual}. After the dual is constructed, the user can decide to redesign, manipulate or optimize the geometry of the dual by assigning different values to the parameters relevant to each methods.



%% file: application.tex
\section{Application}
\label{sec:appl}
The algebraic approach in constructing reciprocal diagrams has three main applications in the context of 3DGS: funicular form finding, structural analysis, and constrained polyhedral manipulations. The following sections will expand on each application.

\subsubsection*{Compression/tension--only form finding}

The algebraic approach enables us to explore a variety of spatial configuration of the forces as funicular forms with compression/tension--only internal forces as well as structural forms with both tensile and compressive internal forces. In the form-finding application, the input is the force diagram, and the user should choose the GFP to specify the direction of the NFPs.

Consider a primal complex which includes closed, convex polyhedral cells. If the external polyhedral cell is chosen as the GFP, with the direction of its faces towards inward, then the dual with all positive edge lengths will represent the equilibrium of a compression-only dual/funicular form.

Figures \ref{fig:iass_01},\ref{fig:dof1},\ref{fig:dof2},\ref{fig:dof3}a,b show the force polyhedron with convex cells as the primal and their compression--only forms as a result of using algebraic method. Note that in all these examples, the GFP is chosen as the exterior polyhedron in the primal.

\subsubsection*{Compression and tension combined form finding}
As mentioned in Section \ref{sec:prime_force}, for a given primal as the force diagram with closed polyhedral cells, choosing any cell other than external force polyhedron results in a structural form with the compression and tension combined internal forces (Figure \ref{fig:tent-comp}).

\subsection*{Constrained polyhedral manipulation of the form}
Often, a designer needs to change/manipulate the geometry of the dual or form of the structure to address certain boundary conditions or to change the location of the applied loads. Algebraic computation of the dual allows for manipulating the geometry of the dual without breaking the reciprocity between two diagrams, i.e., without changing the direction of the members and preserving the planarity of the faces of the dual. 

As mentioned in Section \ref{sec:dsi}, the dimension of the right nullspace of the equilibrium matrix $\textbf{A}$ is the geometric degrees of (in)determinacy of the dual. If the degree is larger than one, there are multiple solutions with significantly different edge lengths and geometrically different forms. 

\begin{figure}[ht!]
  \includegraphics[width=\columnwidth]{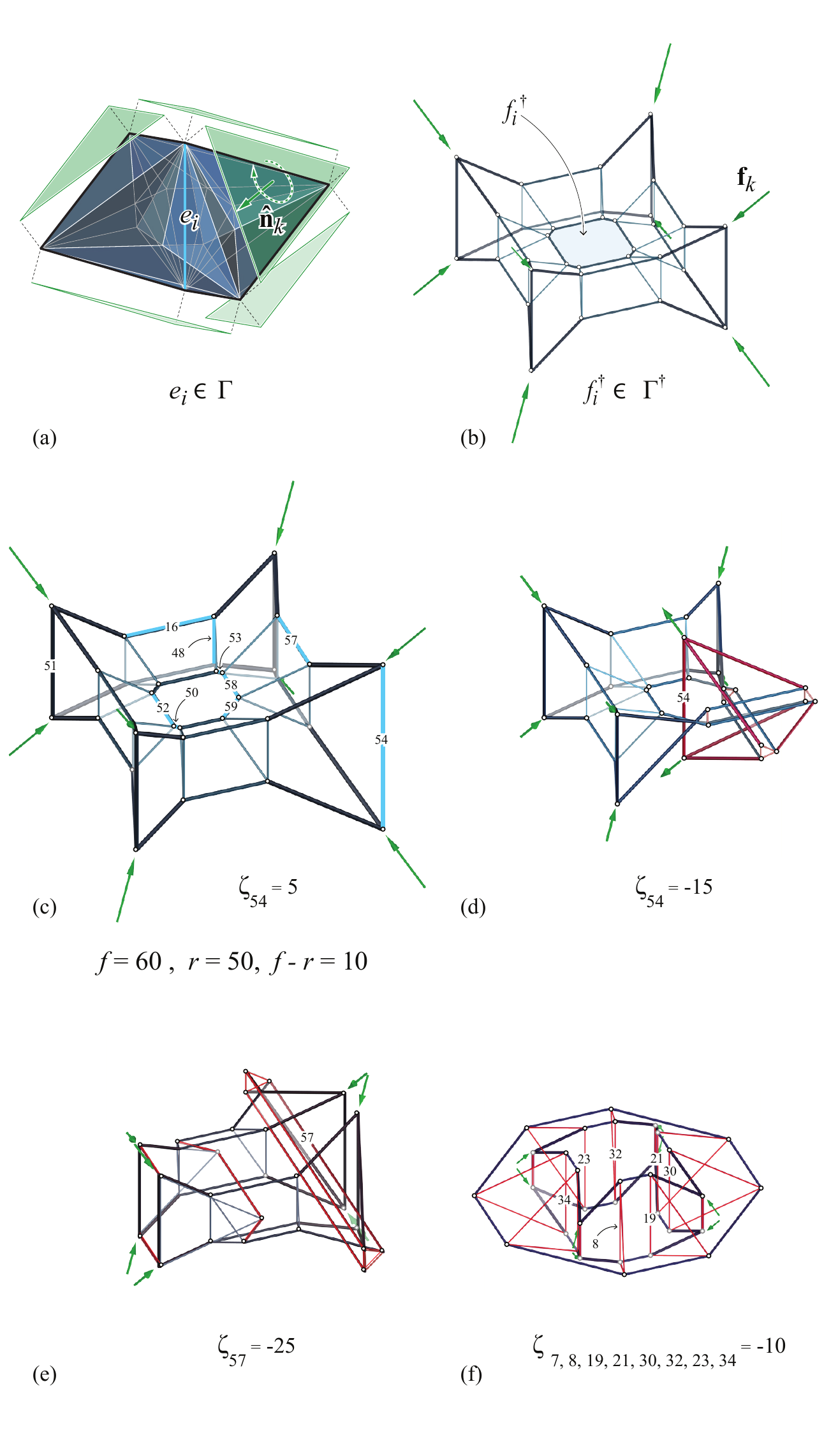}
  \caption{A force diagram as a primal with the external cell as its GPF (a) and the reciprocal diagram computed by using algebraic methods (b) that has 10 degrees of geometric indeterminacy highlighted as the independent edges (c) and the user input parameters to explore variety of compression--and--tension combined forms in equilibrium (d), (e) and (f).     
  }
  \label{fig:dof1}
\end{figure}

For instance, Figure \ref{fig:dof1}b has $10$ degrees of indeterminacy, and the user can explore a variety of solutions by changing the length of the independent edges of the dual. The independent edges can be identified using the RREF method as explained in Section \ref{sec:rref}. The user can specify the lengths of these edges by assigning positive or negative values to the corresponding coordinates of $\zeta$ and recompute the dual with the change in its geometry (Figure \ref{fig:dof1}c--f).

\begin{figure*}[ht]
  \makebox[\textwidth][c]{
  \includegraphics[width= .9\textwidth]{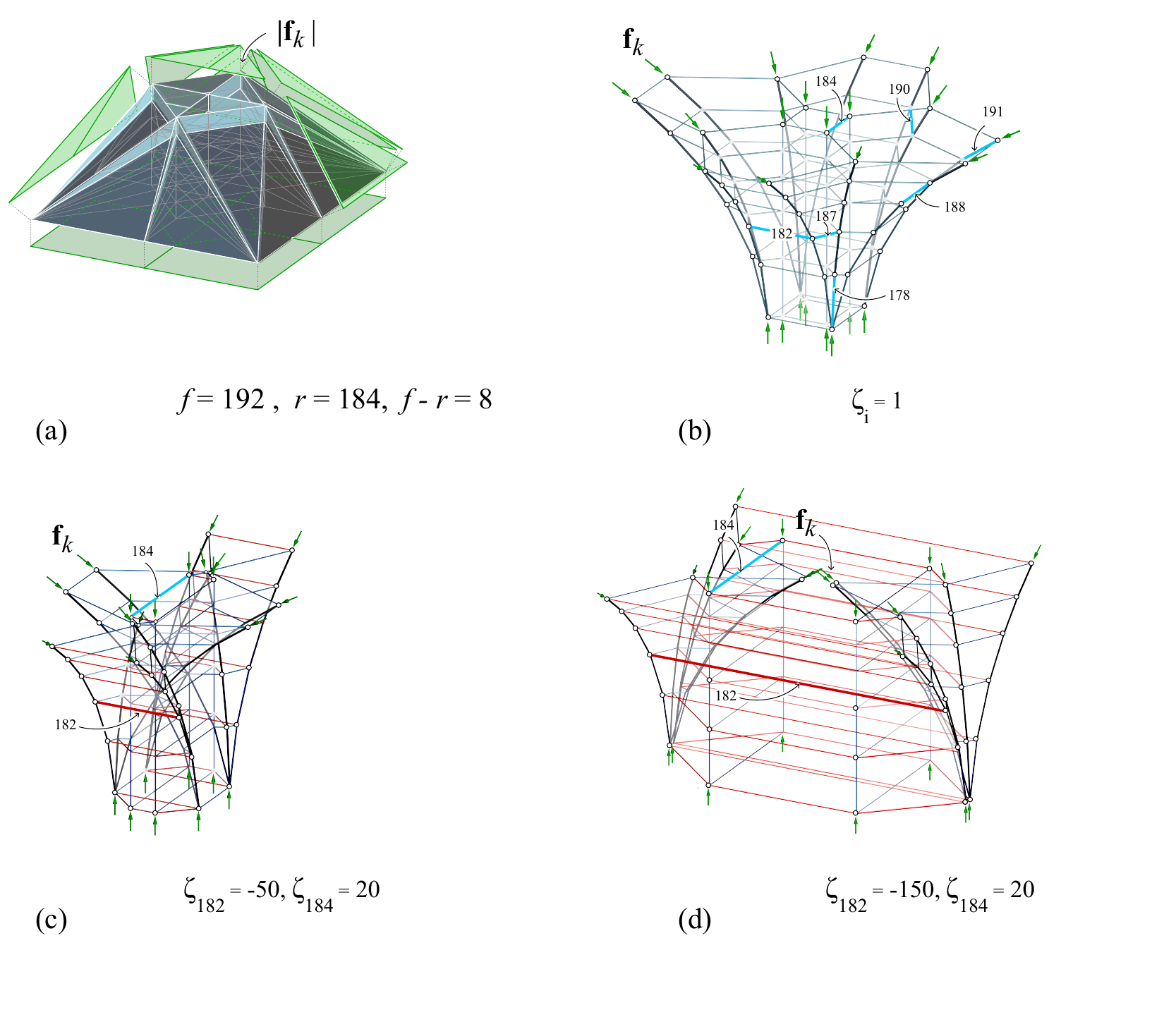}
  }
  \caption{A force diagram as a primal and its dual with 8 degrees of geometric indeterminacy (a) and the compression--only as well as compression--and--tension combined reciprocal diagrams computed by applying the user input parameters (b), (c) and (d).
  }
  \label{fig:dof2}
\end{figure*}

\begin{figure*}[ht!]
 \makebox[\textwidth][c]{
  \includegraphics[width= .9\textwidth]{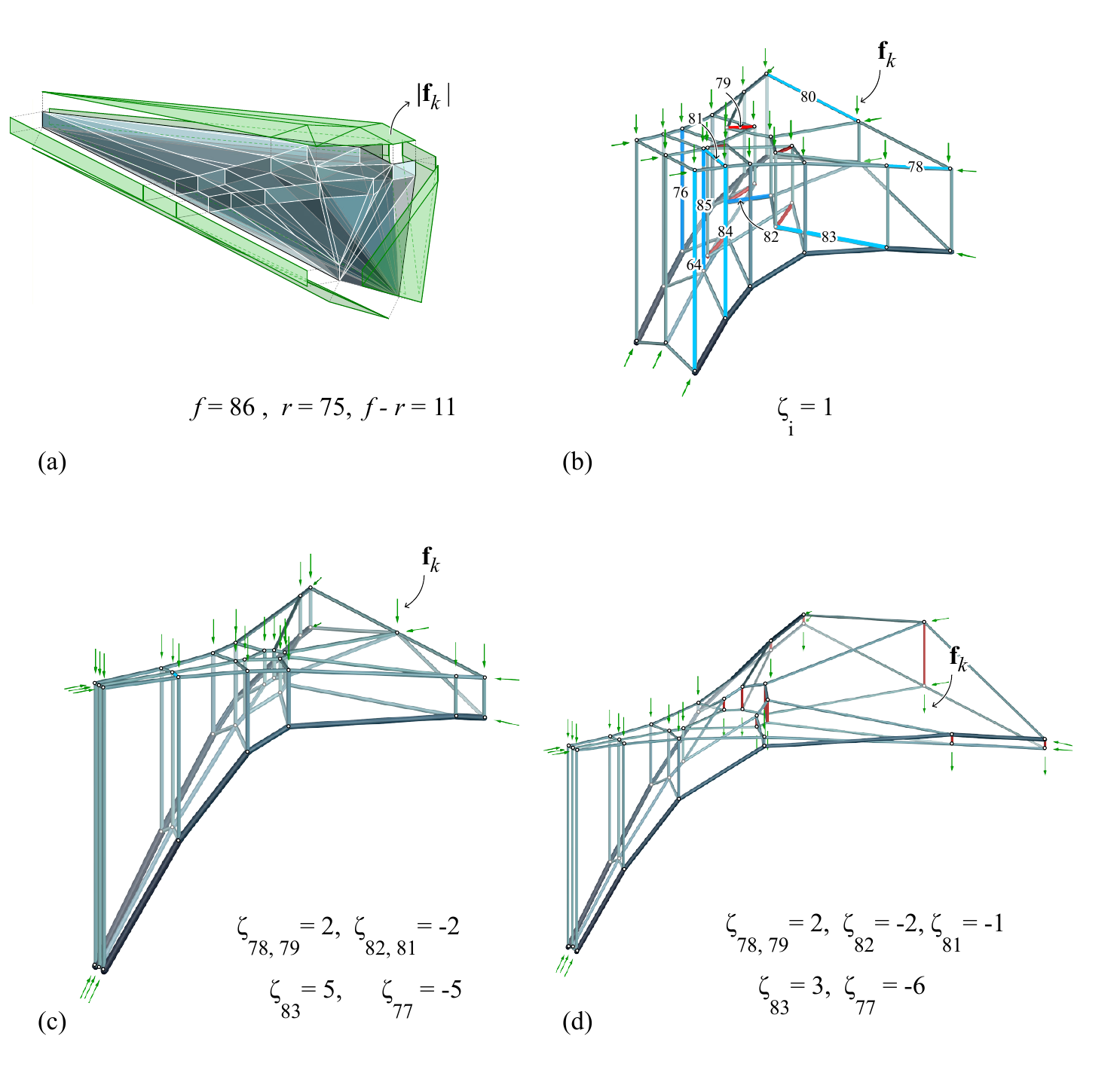}
  }
  \caption{
A force diagram as a primal and its dual with 11 degrees of geometric indeterminacy (a) and  various compression--and--tension combined reciprocal diagrams computed by applying the user input parameters (b), (c) and (d).
  }

  \label{fig:dof3}
\end{figure*}

In Figures \ref{fig:dof1}b,c, the values of $\textbf{q}$ are all positive which results in a compression-only solution; whereas in Figures \ref{fig:dof1}d--f, the values are a combination of positive and negative resulting in systems with combinations of both tensile and compressive forces for the same input force diagram. Figures \ref{fig:dof2} and \ref{fig:dof3} also show the method used to calculate the dual from an input force diagram where the user changes the values of $\zeta$ and calculates various family of solutions with both tensile and compressive internal forces. Therefore, the algebraic method allows us to explore a variety of spatial structural forms with both tensile and compressive forces without changing the force equilibrium.

\subsection*{Structural analysis}
The method explained in this paper can be used for both form finding and analysis as described in Sections \ref{sec:prime_force} and \ref{sec:primal_as_form}. If the primal is considered the form, the dual will represent its force diagram. For statically determinate cases, there will always be a single solution (up to translation and scaling). Therefore, all the examples used in previous sections can be used in a reverse order as shown in Figure \ref{fig:iass_01}. 

For indeterminate cases, the method can be used to explore variety of equilibrium states with various internal and external forces. Although for statically indeterminate cases, we might be able to change the edge lengths of the dual which is the force diagram, controlling the area of the faces and optimizing the values of the internal and external forces of the dual requires additional set of constraints that were not addressed in this paper and will be investigated in future research.

%% file: conclusion.tex
\section{Conclusions and discussions}
\label{sec:conc_disc}
This paper provided an algebraic formulation to construct reciprocal polyhedral diagrams of 3D graphic statics. The approach can be used to construct both form and force diagrams based on the interpretation of the input. The paper explained the process of developing the algebraic constraints and the equilibrium equations for the reciprocal diagrams and provided three computational methods including the Moore-Penrose inverse method (MPI), the Reduced Row Echelon Form (RREF) approach and the Linear programming method (LP) to solve the equilibrium equations. 

The MPI method can be used to construct symmetrical reciprocal diagrams if the primal is symmetrical; the RREF approach can be used to identify the independent edges of the dual that allows generating various solutions with different edge lengths and proportions. The LP method is an excellent approach to generate compression-only results since both MPI and RREF might result in the dual with positive and negative edge lengths. 

Additionally, the paper provides insights in determining the geometric/static degrees of (in)determinacy of the reciprocal diagrams. For indeterminate cases, the deliberate control of the edge lengths allows exploring and manipulating a variety of solutions in equilibrium without changing the planarity of the faces and breaking the reciprocity between two diagrams which is a significant achievement of this paper.    

\subsection*{Limitations and future research directions}
The current approach has the following limitations; although the dual can be a group of polyhedrons with complex (self-intersecting) faces, the current implementation does not accept input with complex (self-intersecting) faces. Expanding the functionality of the data structure to work with self-intersecting cells as input will improve the functionality of the computational workflow that will certainly be addressed in future research.

The algebraic formulation of this paper is capable of constructing a reciprocal force diagram for determinate form diagrams which is unique (up to translation and scaling), and the areas of the faces represent the magnitude of the forces in the primal.     
Although in graphic statics usually designers deal with the statically determinate structural system, controlling the areas of the faces of the dual for indeterminate primal/forms was not addressed in this paper. 

In indeterminate cases, there are multiple force distributions to describe the equilibrium of the form, and controlling the areas of the faces of the dual allows to find the optimized solution among them. Constructing optimized reciprocal constructions by controlling the areas of the faces using algebraic approach is the next step of this research.

The current computational methods heavily rely on precise calculation of the rank of the equilibrium matrix. In other words, the geometric degrees of (in)determinacy of the dual complex is determined by the rank ($r$) of the equilibrium matrix $\textbf{A}$. Thus, small accumulation of numerical errors might result in an imprecise calculation of $r$ that, in turn, leads to an incorrect dual complex.